\documentclass{article}

\usepackage[english]{babel}

\usepackage[a4paper,top=2cm,bottom=2cm,left=3cm,right=3cm,marginparwidth=1.75cm]{geometry}

\usepackage[bookmarks=true,hypertexnames=false,pagebackref]{hyperref}
\hypersetup{colorlinks=true, citecolor=blue, linkcolor=red, urlcolor=blue}
\usepackage[textsize=tiny]{todonotes}

\usepackage{mathrsfs}
\usepackage{graphicx}

\usepackage{amsthm,amsmath,amssymb}
\usepackage{xifthen}
\usepackage[ruled,linesnumbered,vlined]{algorithm2e}

\usepackage{cleveref} 
\usepackage{cite}

\newtheorem{theorem}{Theorem}[section]
\newtheorem{corollary}[theorem]{Corollary}
\newtheorem{lemma}[theorem]{Lemma}

\newtheorem{definition}[theorem]{Definition}

\def\*#1{\mathbf{#1}} \def\+#1{\mathcal{#1}} \def\-#1{\mathrm{#1}}\def\!#1{\mathtt{#1}}\def\@#1{\mathscr{#1}}

\renewcommand{\Pr}[2][]{ \ifthenelse{\isempty{#1}}
  {\mathbf{Pr}\left[#2\right]} {\mathbf{Pr}_{#1}\left[#2\right]} }
\newcommand{\E}[2][]{ \ifthenelse{\isempty{#1}}
  {\mathbf{E}\left[#2\right]}
  {\mathbf{E}_{#1}\left[#2\right]} }
\newcommand{\Var}[2][]{ \ifthenelse{\isempty{#1}}
  {\mathbf{Var}\left[#2\right]}
  {\mathbf{Var}_{#1}\left[#2\right]} }
\newcommand{\Ent}[2][]{ \ifthenelse{\isempty{#1}}
  {\mathbf{Ent}\left[#2\right]}
  {\mathbf{Ent}_{#1}\left[#2\right]} }

\renewcommand{\emptyset}{\varnothing}

\allowdisplaybreaks

\newcommand*\samethanks[1][\value{footnote}]{\footnotemark[#1]}

\title{MMS Allocations of Chores with Connectivity Constraints: New Methods and New Results}
\author{Mingyu Xiao
	\thanks{University of Electronic Science
and Technology of China.\{myxiao,huangsen47\}@gmail.com.}
        \and Guoliang Qiu \thanks{Shanghai Jiao Tong University. guoliang.qiu@sjtu.edu.cn.}
	\and Sen Huang\samethanks[1]}

\begin{document}
\maketitle

\begin{abstract}
    We study the problem of allocating indivisible chores to agents under the Maximin share (MMS) fairness notion. The chores are embedded on a graph and each bundle of chores assigned to an agent should be connected. Although there is a simple algorithm for MMS allocations of goods on trees, it remains open whether MMS allocations of chores on trees always exist or not, which is a simple but annoying problem in chores allocation.  In this paper, we introduce a new method for chores allocation with connectivity constraints, called the group-satisfied method, that can show the existence of MMS allocations of chores on several subclasses of trees. Even these subcases are non-trivial and our results can be considered as a significant step to the open problem. We also consider MMS allocations of chores on cycles where we get the tight approximation ratio for three agents. Our result was obtained via the linear programming (LP) method, which enables us not only to compute approximate MMS allocations but also to construct tight examples of the nonexistence of MMS allocations without complicated combinatorial analysis. These two proposed methods, the group-satisfied method and the LP method, have the potential to solve more related problems.
\end{abstract}

\section{Introduction}\label{sec:intro}

The theory of the \emph{fair allocation} investigates how to allocate items to a set of agents under some fairness notion, where different agents may have different valuation functions on the items. The spirit of the fair allocation problem is to achieve a desired outcome for individuals and the whole community simultaneously, which motivates several important problems in mathematics and computer science. The \emph{goods allocation} problem with positive valuation functions has received extensive studies~\cite{LMMS04,budish2011,BL16,markakis2017}.
However, in some scenarios in real life, the items to be allocated may have disutility, i.e., the valuation functions are negative, such as troublesome jobs, household chores, or debt. For this case, the problem is called the \emph{chores allocation} problem~\cite{ARSW17}.
Seemingly, the problem of chores allocation is similar to the well-studied goods allocation problem as we can reduce the former to the latter by setting all valuation functions as their negative counterparts.  However, many properties of the problem may change and not be applicable under this reduction. Thus, most results for goods allocation cannot be trivially extended to chores allocation~\cite{ARSW17,BK20,HL21,ALW19,ZW22}.


There are several fairness notions for allocations, such as envy-free (EF)~\cite{LMMS04}, proportionality (PROP)~\cite{steihaus1948}, maximin share (MMS)~\cite{budish2011}, and so on.
In this paper, we consider the MMS fairness notion where the MMS value is the best possible guarantee that an agent can expect when he divides the items into several bundles but picks after all other agents. Moreover, we study MMS allocations of chores on graphs with connectivity constraints in which chores are embedded on a graph and each bundle assigned to an agent is required to be connected~\cite{DBLP:journals/corr/abs-1808-09406,DBLP:journals/corr/abs-1811-04872,BCEIP17,LT18,BCL19}. The connectivity requirements for chores capture the scenarios such as the allocation for energy conservation and emission reduction obligations between countries where the feature of geographical adjacency is natural, crop harvesting task allocation problem or clean work arrangement in the street where arranging a task with continuous geographical location is more convenient to set and use tools, and so on.
Note that MMS allocation may not always exist. It is also meaningful to study the approximate $\alpha$-MMS allocation: whether each agent can always get $\alpha$ fraction of his MMS guarantee in the allocation.

\subsection{Our Contribution}

We propose two novel techniques to study MMS allocation of chores with connectivity constraints and demonstrate their applications
on trees and cycles. Although the graph classes are restrictive, claiming the existence of the (approximate) MMS allocations of chores with these constraints is already non-trivial and requires sophisticated analysis.

For trees, the simple algorithm for MMS allocation of goods~\cite{BCEIP17} can not be extended to chores and an illustrated example will be presented in our later discussion. Resulted from the subtle difference between the goods and chores settings, it remains unknown whether or not MMS allocation of chores on a tree always exists.
We generalize the classical last-diminisher method to a method called the group-satisfied method. Together with a matching technique in graph theory, we show the existence of MMS allocations of chores on some special trees, such as trees with depth at most 3, spiders, and so on.

For cycles, the idea for goods allocation in \cite{LT18} can be trivially used to obtain a 3/2-MMS allocation of chores on cycles. Here our major contribution is the idea of using linear programming to design algorithms for approximate $\alpha$-MMS allocations and construct examples to show the nonexistence of MMS allocations. We take the problem of allocating chores on a cycle to three agents as an example to show the linear programming method.
By using the linear programming method, we may be able to avoid some complicated combinatorial analysis. We show a tight result: $7/6$-MMS allocation of chores on a cycle to three agents always exists and can be found in polynomial time; $\alpha$-MMS allocation may not exist for any constant $\alpha < {7/6}$. The linear programming method may be used to solve more allocation problems. A directed extension to MMS allocation of goods on a cycle to three agents, we also get a tight ratio of $5/6$ for $\alpha$.

In the following part of the paper, we first introduce related work on MMS and preliminary background.
Second, we discuss the difference between allocating goods and chores on trees and introduce the group-satisfied method.
Third, we show how to use the group-satisfied method to find MMS allocations for chores on two kinds of trees: trees with depth 3 and spiders.
Fourth, we consider MMS allocations of chores on cycles and introduce a linear programming method to find approximation MMS allocations.
Finally, we discuss the feasibility of our new methods for other problems.


\subsection{Other Related Works}

For goods allocation, the MMS fairness notion was first introduced in~\cite{budish2011}.
The first instance where MMS allocation does not exist was given in~\cite{KPW18}.  An instance with a fewer number of goods was identified in~\cite{kPW16}. On the other hand, a 2/3-MMS allocation always exists~\cite{KPW18}. Later, the ratio was improved to 3/4~\cite{GHSSY18} and to $3/4+1/(12n)$~\cite{GT20}, where $n$ is the number of agents.
For only three agents and four agents, the ratio can be further improved to 7/8~\cite{AMNS17} and 4/5~\cite{GHSSY18} respectively.
It is also known that a 39/40-MMS allocation for three agents is impossible~\cite{FFT21}.
All the above results do not take into account connectivity constraints.
For goods allocation on a graph with connectivity constraints, MMS allocation always exists on trees~\cite{BCEIP17} but may not exist on single cycles~\cite{LT18}.
For goods on a cycle, it is known that 1/2-MMS allocation always exists and the ratio can be improved to 5/6 if there are only three agents~\cite{LT18}.

For the chores setting, MMS allocation may not always exist, but 2-MMS allocation always exists~\cite{ARSW17}.
The result was later improved to $4/3$~\cite{BK20} and to $11/9$ \cite{HL21}.
For chores with connectivity constraints, EF, PROP, and Equitability allocations on paths and stars were studied in~\cite{BCL19}. A more relaxed fairness notion was studied on paths in~\cite{BCFIMPVZ22}.
Nevertheless, the MMS criterion has not been well explored, except for some trivial results extended from goods allocation.

\section{Backgrounds}
Let $A=\{1,\dots,n\}$ be a set of $n$ agents and $C=\{c_1,\dots,c_m\}$ be a set of $m$ chores.
There is an undirected graph $G=(C,E)$ with vertices being chores in $C$.
For each agent $i\in A$, there is a disutility function on chores $u_i:C\rightarrow\mathbb{R}_{\leq 0}$,
where the functions $u_i$ are \emph{additive}, i.e., $u_i(C')=\sum_{c\in C'} u_i(c)$ holds for any subset of chores $C'\subseteq C$.
The whole disutility function profile for all agents is denoted by $U=\{u_1,\dots,u_n\}$.

Let $k$ be a positive integer. Let $P^k: [k] \to 2^C$ be a \emph{$k$-partition} of $C$ such that $\cup_{i\in [k]}P^k(i)=C$ and $P^k(i)\cap P^k(j)=\emptyset$ for any different $i,j\in [k]$, where $[k]=\{1,\dots,k\}$ and the set $P^k(i)$ is called the $i$th \emph{bundle} in the $k$-partition.
A $k$-partition $P^k$ is \emph{valid} if the induced subgraph $G[P^k(i)]$ is connected for each $i\in [k]$.
Let $\mathcal{P}^k$ denote the set of all valid $k$-partitions.
An \emph{allocation} of $C$ is defined as a valid $n$-partition $\pi:A\rightarrow 2^C$.

We mainly consider the \textit{Maximin share (MMS)} criterion for the allocation of chores.
Given a graph $G$ and a positive integer $k$, for each $i\in A$, we define the \emph{MMS value} of agent $i$ as follows when the chores in $G$ are allocated to $k$ agents
$$mms_i^k(G)=\max_{P^k\in\mathcal{P}^k}\min_{j\in [k]}u_i(P^k(j)).$$
For simplification, we use $mms_i^k$ and $mms_i$ to denote $mms_i^k(G)$ and $mms_i^n$, respectively.

For any constant $\alpha \geq 1$.
An $n$-partition $P$ of $G$ is called an \emph{$\alpha$-MMS$_{i}$ split} for agent $i$ if each bundle $b\in P$ induces a connected subgraph and $u_i(b)\geq \alpha\times mms_i$.
When $\alpha=1$, we simply call it an MMS$_{i}$ split.
We say a valid allocation $\pi$ is an \emph{$\alpha$-MMS allocation} if $u_i(\pi(i))\geq \alpha\times mms_i$ for each agent $i\in A$. When $\alpha=1$, we simply call it an MMS allocation.

We focus on the existence of MMS (or $\alpha$-MMS) allocations of chores on graphs and algorithms to find them if they exist. Our algorithms may need to use the MMS value for each agent.
We show that when the graph is a tree or a cycle, the MMS value for each agent can be easily computed. The algorithms are modified from the algorithms for goods in~\cite{BCEIP17} and the detailed proof can be found in 
Appendix~\ref{append-1}.

\begin{lemma} \label{MMScom}
For allocating chores on a tree or a cycle, the MMS value $mms_i$ for each agent $i\in A$ can be computed in polynomial time.
\end{lemma}

\section{Failure of the Last-diminisher Method on Trees for Chores Allocations}
As mentioned in the introduction, MMS allocations of goods on trees always exist and can be computed in polynomial time~\cite{BCEIP17}.
The algorithm uses the idea of the last-diminisher method for proportionality cuttings of divisible cake~\cite{cake16}.

We first review the allocation procedure for goods. First,  let one agent take a sub rooted tree as $T$ (the original tree is rooted by taking an arbitrary vertex as the root);
second, each other agent, in order, replaces $T$ with a sub rooted tree $T'$ of it if he thinks $T$ is too much and does nothing if he thinks the current subtree $T$ is not enough or just right;
third, the last agent who modifies $T$ gets the bundle $T$. Note that after deleting $T$ from the original tree $G$, the remaining part is still a connected tree.
The same procedure is then applied to allocate the remaining tree to the remaining agents. This algorithm is not exactly the algorithm for allocating goods on trees in~\cite{BCEIP17}.
Both algorithms use the idea of the last-diminisher method and we just present a simplified version.

\begin{figure}[h]
    \centering
        \includegraphics[width=4cm,height=2.1cm]{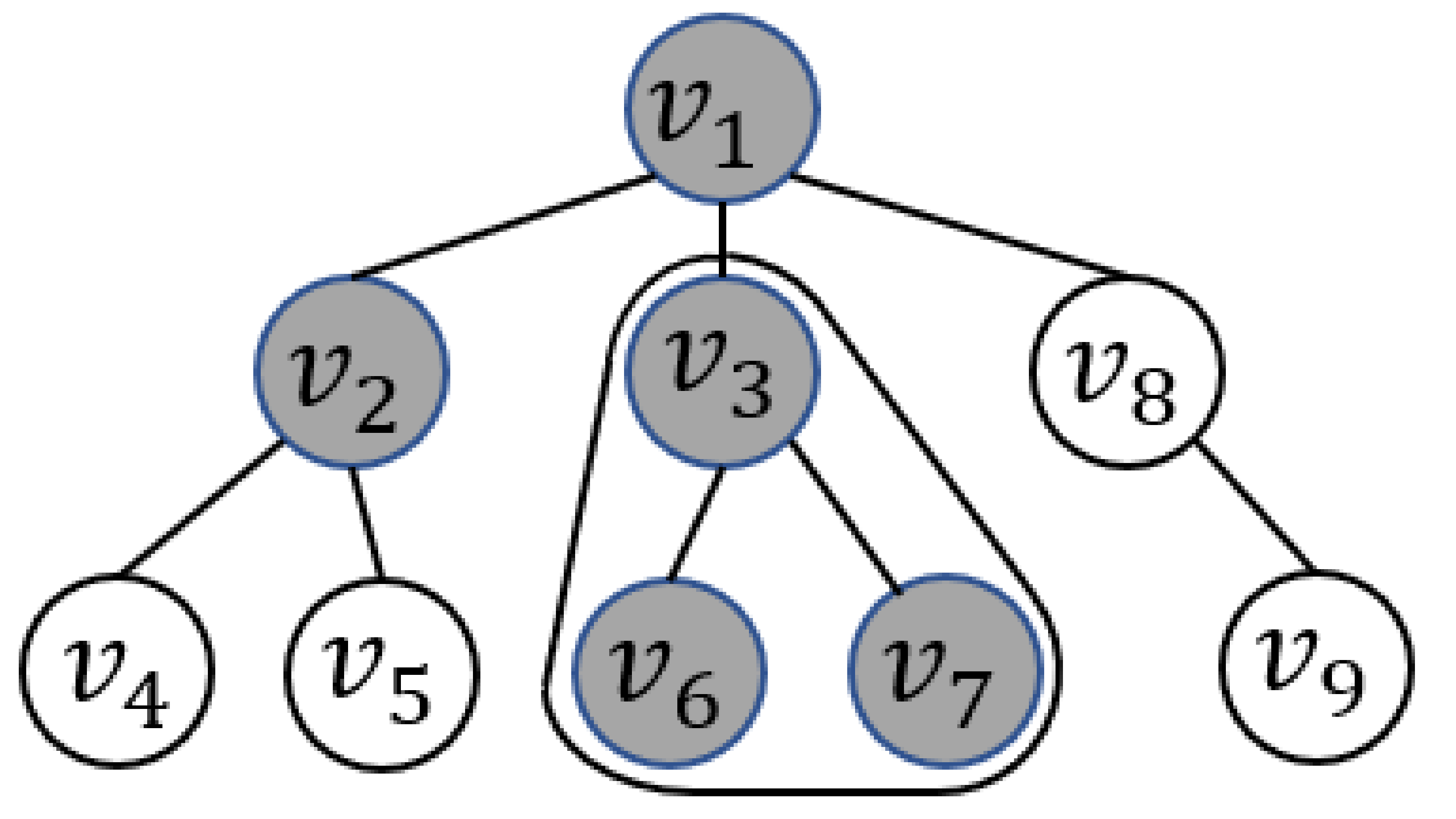}
    \caption{The initial bundle contains $v_3, v_6$ and $v_7$. After adding $v_1$ and $v_2$ in the bundle, the new bundle is not a rooted subtree anymore.}
    \label{hard}
\end{figure}

For chores, agents do not want to get more burdens. The corresponding last-diminisher method will be different: each agent may want to expand the current bundle if he thinks the current bundle of burdens is too light.
At first glance, we can also assign the last expanded bundle to the last agent who expands it.
However, the expanding operation will cause trouble on trees. After expanding the current bundle (a subtree) by including some vertices to it, the remaining graph after deleting the new bundle may not be connected (not be a connected tree anymore). This may change the property of the problem. Specifically, we give an example in Figure~\ref{hard} to show that after expanding a rooted subtree the remaining part may not be connected. The current subtree $T$ is the subtree rooted as $v_3$, which contains three vertices $\{v_3, v_6,v_7\}$. An agent may include two vertices $v_1$ and $v_2$ to the subtree $T$. However,
after adding them, $T$ is not a rooted subtree and the remaining graph after deleting $T$ is not connected.
For this case, we can not guarantee the existence of allocations to the remaining $n-1$ agents even if the original allocations to the $n$ agents exist.

It turns out that chores allocation on trees become much harder. We show that chores allocation on some special trees always exists and can be found in polynomial time.
Our algorithms will use a technique, called the \emph{group-satisfied method}, which can be regarded as an extension of the last-diminisher method. In an allocation procedure of the group-satisfied method, we will assign $k'$ bundles to $k'$ agents together such that the remaining $n-k'$ agents `agree with' this allocation and the remaining objects still form a connected tree. Here `agree' means that the agent thinks in the new instance his MMS value will not decrease. The group-satisfied method will iteratively apply this allocation procedure until all objects are assigned to agents.
If in each allocation procedure only one bundle is assigned to one agent, then the group-satisfied method becomes the last-diminisher method.
Before introducing the group-satisfied method in detail, we first define some concepts.

\section{Preliminaries for the group-satisfied method}
Recall that a partial allocation is valid if each bundle of chores in the allocation induces a connected graph.

\begin{definition}
Let $\pi'$ be a valid partial allocation of a subset set of chores $C'\subseteq C$ to a subset of agents $A'\subseteq A$.
An agent $i\in A'$ is \emph{satisfied} with $\pi'$ if the disutility of the bundle assigned to him is not less than his MMS value, i.e., $u_i (\pi'(i))\geq mms_i$;
an agent $i\in A\setminus A'$ is \emph{satisfied} with $\pi'$ if in the remaining instance his MMS value is not decreasing, i.e., $mms_i^{n-|A'|}(G[C\setminus C'])\geq mms_i^{n}(G)$ for each agent $i\in A\setminus A'$.
\end{definition}

\begin{definition}
A valid partial allocation $\pi'$ is \emph{group-satisfied} if all agents are satisfied with it.
\end{definition}

Given a  valid partial allocation $\pi': A'\rightarrow 2^{C'}$. To check whether $\pi'$ is group-satisfied we need to check whether all agents are satisfied with it.
For an agent in $A'$, it is easy to check since we only need to compute the disutility of the bundle assigned to him. For an agent $i\in A\setminus A'$, we use the following sufficient condition to guarantee their requirements,
\begin{lemma} \label{forsaft}
Let $\pi': A'\rightarrow 2^{C'}$ be a valid partial allocation.
Let $i$ be an agent in $A\setminus A'$ and $P_i$ be an MMS$_i$ split of $G$.
If $G[C\setminus C']$ contains chores only from at most $n-|A'|$ bundles of $P_i$ and the chores from the same bundle induce a connected graph in $G[C\setminus C']$, then $i$ is satisfied with $\pi'$.
\end{lemma}
\begin{proof}
    For agent $i\in A\setminus A'$, we split $G[C\setminus C']$ into at most $n-|A'|$ bundles according to the MMS$_i$ split $P_i$ of $G$, i.e., chores in the same bundle of $P_i$ will still be in the same bundle.
    We can see  that each bundle is also connected. Furthermore,  the disutility of each bundle is not less than $mms_i^{n}(G)$. This split is a valid $n-|A'|$-partition of $G[C\setminus C']$. So we know that $mms_i^{n-|A'|}(G[C\setminus C'])\geq mms_i^{n}(G)$.
\end{proof}

A subclass of trees satisfying a certain property $\Pi$ is called \emph{Property-$\Pi$ graphs}. For example, the class of paths is trees having the property: having no vertices of degree $\geq 3$.
In this paper, we will consider several Property-$\Pi$ graphs, such as paths, trees with only one vertex of degree $\geq 3$, trees with depth 3, and so on.
We have the following important lemma that will widely be used in our proofs.

\begin{lemma} \label{important}
For any instance of MMS allocation of chores on Property-$\Pi$ graphs, if there is always a group-satisfied partial allocation $\pi': A'\rightarrow 2^{C'}$ such that the remaining graph $G[C\setminus C']$ is still a Property-$\Pi$ graph,
then MMS allocations of chores on Property-$\Pi$ graphs always exist. Furthermore, if the partial allocation $\pi'$ can be found in polynomial time, then MMS allocations of chores on Property-$\Pi$ graphs can be found in polynomial time.
\end{lemma}

\begin{proof}
    We prove this lemma by induction on the number of agents. It trivially holds for one agent: MMS allocations always exist and they can be found in polynomial time. Next, we assume that the lemma holds for any $n'<n$ agents and proves that the lemma also holds for $n$ agents. Let  $\pi': A'\rightarrow 2^{C'}$ be a group-satisfied partial allocation and there is an algorithm that can find it in polynomial time.
    We consider the remaining problem after the allocation $\pi'$. The remaining set of chores $C''=C\setminus C'$ also induces a graph $G''$ that is a Property-$\Pi$ graph since $\pi'$ is group-satisfied. We need to allocation $C''$ to the remaining $n-|A'|$ agents in $A''=A\setminus A'$.
    By the induction, we know that there is an MMS allocation $\pi'': A''\rightarrow 2^{C''}$ such that $u_i (\pi''(i))\geq mms_i^{|A''|}(G'')$ holds for each $i\in A''$.
    By the definition of  group-satisfied allocations, we also know that $mms_i^{|A''|}(G'')\geq mms_i^n(G)$ holds for each $i\in A''$.
    Therefore, the two allocations $\pi'$ and $\pi''$ will form a valid MMS allocation of the original graph $G$.

    For the running time, by the induction, we know that the allocation $\pi''$ can be found in polynomial time. The allocation $\pi'$ can also
    be executed in polynomial time by the assumption. So an MMS allocation of the whole graph $G$ can be done in polynomial time.
\end{proof}

Next, we show how to use Lemma~\ref{important} to find feasible allocations. We first consider the simple case of paths.

\subsection{Paths}
Assume that the graph $G$ is a path now. We consider the MMS$_i$ split of the path $G$ for each agent $i$. In each MMS$_i$ split, the bundle containing the most left chore in the path $G$ is called the \emph{first bundle}.
Since the graph is a path, we know that there is always an agent $i^*$ whose first bundle $C^*$ contains any other agent's first bundle. The allocation procedure is to allocate bundle $C^*$ to agent $i^*$.
We can see that this allocation is group-satisfied and the remaining chores still form a path. By Lemma~\ref{important}, we know that MMS allocations of chores on paths always exist.
For the running time, if we already know the MMS value of each agent, then the algorithm takes only $O(nm)$ time because we check from the left to right at each chore whether the sum of the values of all the chores on the left is greater than the MMS value for each agent.
Note that, in this allocation procedure, only one bundle of chores is assigned to one agent in each iteration. So it is indeed the last-diminisher method.

Before introducing the algorithms for trees of a more complicated structure, 
we introduce more notations.

\subsection{$x$-perfect and $x$-super subtrees}
Next, we assume the input graph $G$ is a tree. We will select a vertex $r$ in $G$ as the root and consider the tree as a rooted tree. The subtree rooted as a vertex $v$ is denoted by $T_v$. The length of the path from the root $r$ to a vertex $v$ in the tree is the \emph{depth} of the vertex $v$. The \emph{depth} of a rooted tree is the largest depth of all vertices.
We will also say an unrooted tree has the depth at most $x$ if we can select a vertex as the root such that the rooted tree has the depth at most $x$.
 A tree of a single vertex is called a \emph{trivial tree} and a tree of more than one vertex is called a \emph{non-trivial tree}.

\begin{figure}[htbp]

    \centering
        \includegraphics[width=4cm,height=2.2cm]{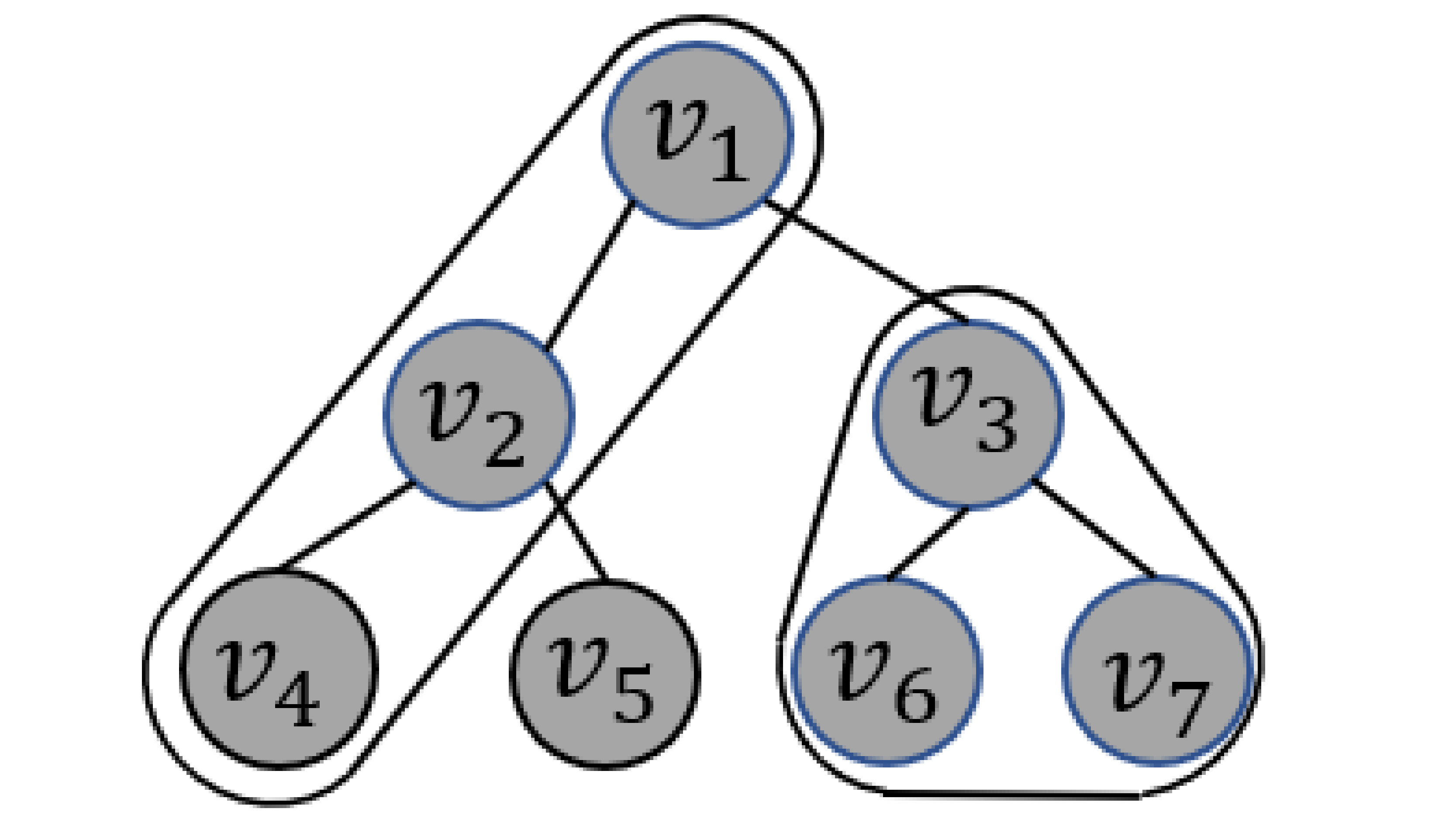}
    \caption{In this instance, there are 7 chores on the tree. One agent's MMS split partitions the tree into three bundles as shown.
    We can see that this agent $2^{+}$-splits the subtree rooted as $v_2$ and $1$-splits the subtree rooted as $v_3$.}
    \label{pv1}
\end{figure}

 For each agent $i\in A$, we fix an MMS$_i$ split $P_i$ of the tree.
Let $v$ be a vertex on the tree. We consider the subtree $T_v$ rooted at $v$. Assume that $x$ bundles in an MMS$_i$ split $P_i$ are contained in the subtree $T_v$, and $y$ bundles in $P_i$ contain at least one vertex in $T_v$.
It always holds that $x\leq y \leq x+1$ since there is at most one bundle in $P_i$, which contains the vertex $v$, that contains some vertices in $T_v$ but is not contained in $T_v$.
We say that agent $i$ \emph{$y$-splits} the subtree $T_v$ if the MMS$_i$ split $P_i$ makes $x=y$
and  \emph{$y^+$-splits} the subtree $T_v$ if the MMS$_i$ split $P_i$ makes $y=x+1$.
See Figure~\ref{pv1} for an illustration of $y$-splitting and $y^+$-splitting.

\vspace{-2mm}
\begin{definition} \label{pands}
A rooted subtree $T_v$ is called \emph{$x$-perfect} if no agent $z$-splits it for any $z< x$, no agent $z^+$-splits it for any $z\leq x$, and there is at least one agent $x$-splits it.
A rooted subtree $T_v$ is called \emph{$x$-super} if no agent $z$-splits it for any $z< x$, no agent $z^+$-splits it for any $z< x$, and there is at least one agent $x^+$-splits it.
\end{definition}

\begin{definition}
For an $x$-super tree $T_v$, an agent that  $x^+$-splits it is called a \emph{dominator} of the tree $T_v$, and the set of dominators of $T_v$ is denoted by $D(T_v)$.
\end{definition}

\begin{lemma}\label{perfect_c}
Let $T_v$ be an $x$-perfect rooted subtree, and $\pi'$ be a valid partial allocation that allocates $T_v$ to $x$ agents $A'$.
If each agent in $A'$ is satisfied with $\pi'$, then $\pi'$ is group-satisfied.
\end{lemma}
\begin{proof}
    We only need to show that any agent in $A\setminus A'$ is also satisfied with $\pi'$. According to the definition of $x$-perfect subtrees, we know that for any agent $i\in A$, the graph $G-T_v$ contains chores from at most $n-x$ bundles of his MMS$_i$ split of $G$. Furthermore, the chores from the same bundle form a connected graph in $G-T_v$ since $T_v$ is a rooted subtree of $G$. By Lemma~\ref{forsaft}, we know that all agents in $A\setminus A'$ are satisfied with $\pi'$.
\end{proof}

\begin{lemma}\label{super_c}
Let $\mathcal{T}=\{T_1, T_2, \dots, T_l\}$ be a set of disjoint rooted subtrees of $G$, where $T_i$ is $x_i$-super $(i\in \{1,2,\dots, l\})$.
Let $\pi'$ be a valid partial allocation that allocates chores in all subtrees in $\mathcal{T}$ to $\sum_{i=1}^l x_i$ agents $A'$, where $A'\supseteq \bigcup_{i=1}^l D(T_i)$ contains all dominators of all subtrees in $\mathcal{T}$.
If each agent in $A'$ is satisfied with $\pi'$, then $\pi'$ is group-satisfied.
\end{lemma}

\begin{proof}
    We only need to show that any agent in $A\setminus A'$ is also satisfied with the allocation $\pi'$.
    For any subtree $T_i\in \mathcal{T}$ and any agent $j\in A\setminus D(T_i)$, by the definition of $x$-super subtrees and dominators, we know that the graph $G-T_i$ contains only chores from at most $n-x_i$ bundles of the MMS$_i$ split of $G$. By iteratively applying this claim, we get that for any agent $j_0 \in A\setminus \Pi_{i=1}^l D(T_i)$, the graph $G-\cup_{i=1}^l T_i$ contains only chores from at most $n- \sum_{i=1}^l x_i$ bundles of his MMS$_{j_0}$ split of $G$. Furthermore, the chores from the same bundle form a connected graph in $G-\cup_{i=1}^l T_i$ since each $T_i$ is a rooted subtree of $G$. We have that $A'\supseteq D(T_v)$.
    By Lemma~\ref{forsaft}, we know that all agents in $A\setminus A'$ are satisfied with $\pi'$.
\end{proof}

Lemmas~\ref{perfect_c} and~\ref{super_c} provide some ideas to construct group-satisfied partial allocations based on $x$-perfect and $x$-super subtrees.
Note that Lemma~\ref{perfect_c} only considers one $x$-perfect subtree while Lemma~\ref{super_c} considers several $x_i$-super subtrees together. For several $x$-perfect subtrees, we can deal with them one by one by using Lemma~\ref{perfect_c}. However, for several $x_i$-super subtrees, we should use the stronger Lemma~\ref{super_c}.

\section{Trees with depth 3}
Now we consider trees with depth at most 3 and show how to use the group-satisfied method to solve the allocation problem on these trees. Let $r$ be the root of the tree $G$. If the depth of the tree is 2, then the tree is a star.
For this case, we consider the MMS$_{i_0}$ split $P_{i_0}$ of an arbitrary agent $i_0$ and let agent $i_0$ take the bundle containing the root. Note that all other $n-1$ bundles in $P_{i_0}$ contain only one chore in one leaf of the tree. We arbitrarily assign the $n-1$ chores to
the other $n-1$ agents, since the value of every single chore will not less than the MMS value of each agent. This will be an MMS allocation. Next, we assume that the depth of $G$ is 3.

In the algorithm, we also first consider the MMS$_{i_0}$ split $P_{i_0}$ of an arbitrary agent $i_0$ and denote the bundle containing the root by $p_r$.
In the graph $G'=G-p_r$ after deleting $p_r$, each connected component is either a star or a single vertex.
Let $C=\{c_1,c_2,\dots,c_p\}$ be the set of stars in $G'$. Each star in $C$ is a rooted subtree of $G$. We will show that we can always find a group-satisfied allocation.

\textbf{Case 1}: There is a star $c_q\in C$ that is $x$-perfect for some integer $x$. Assume that agent $i$ $x$-splits star $c_q$. We consider the $x$-partition of $c_q$ by agent $i$ and assign the bundle containing the center vertex of $c_q$ to agent $i$ and
assign the left $x-1$ chores to arbitrary $x-1$ agents. All the $x$ agents are satisfied with this allocation. By Lemma~\ref{perfect_c}, we know that this allocation of $c_q$ to the $x$ agents is group-satisfied.

\textbf{Case 2}: All stars in $C$ are $x$-super for different integers $x$. Assume that $c_i$ is $x_i$-super for $1\leq i\leq p$. In this case, we will use a matching technique to show that there is either a group-satisfied allocation to allocate some stars in $C$ or a group-satisfied allocation to allocate the whole graph $G$.

\begin{figure}[htbp]

    \centering
        \includegraphics[width=9cm,height=5cm]{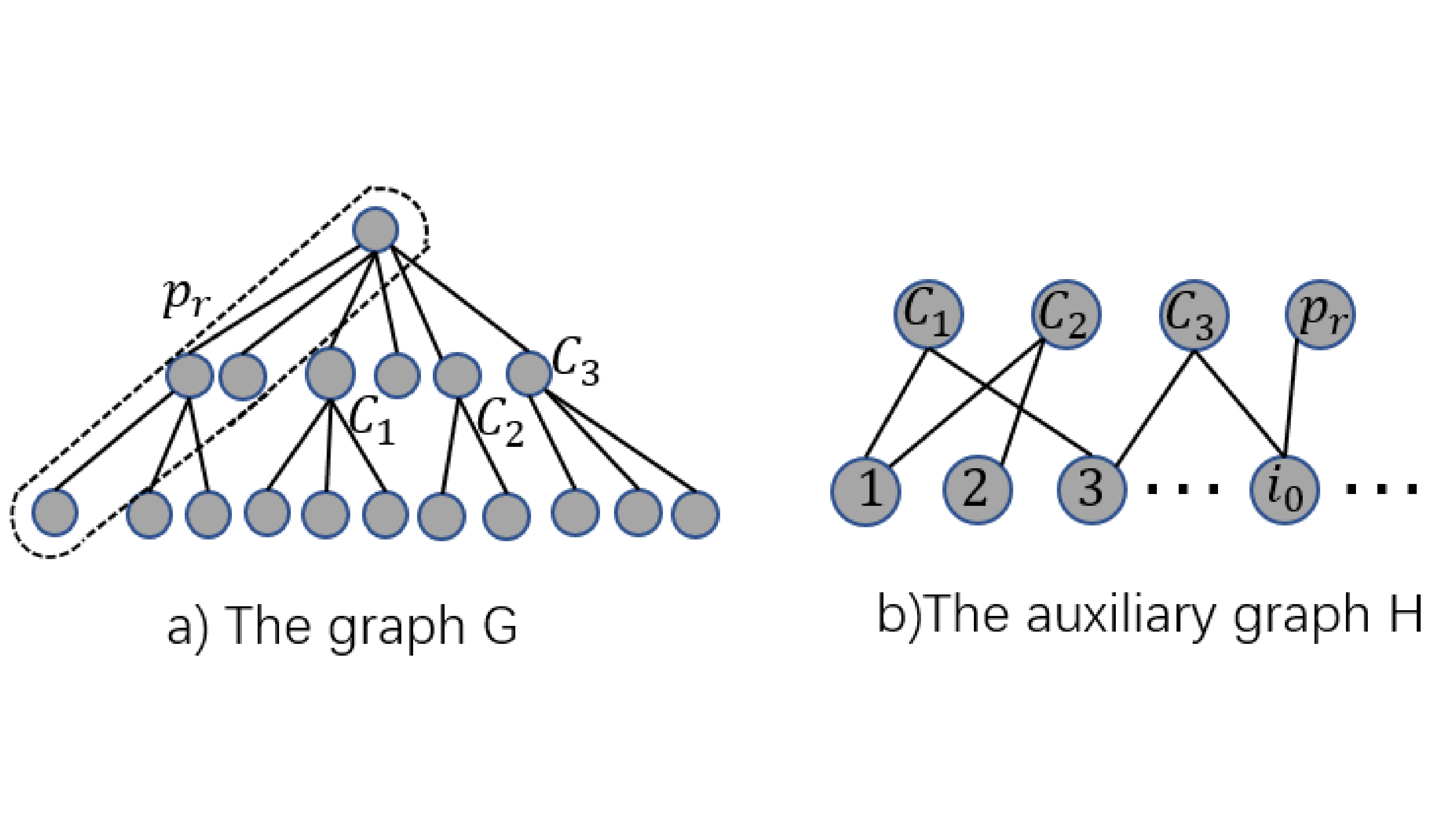}
       \caption{The graph $G$, where $D(C_1)=\{1,3\},D(C_2)=\{1,2\}$ and $D(C_3)=\{3,i_0\}$, and the auxiliary graph $H$.}
    \label{auxgraph}
\end{figure}

We construct an auxiliary bipartite graph $H=(V_1, V_2, E_H)$. The vertex set $V_1$ contains $|C|+1$ vertices. Each star in $C$ is corresponding to a vertex in $V_1$ and the last vertex in $V_1$ is corresponding to the bundle $p_r$ in $P_{i_0}$, i.e., $V_1=C\cup\{p_r\}$. The vertex set $V_2$ is corresponding to the agent set $A$, i.e., $V_2=A$. A vertex in $c_j\in C$ is adjacent to a vertex $i\in V_2$ if and only if the corresponding agent $i$ is a dominator of the subtree $c_j$, i.e., $i\in D(c_j)$. The vertex $p_r\in V_1$ is only adjacent to vertex $i_0\in V_2$. See Figure~\ref{auxgraph} for an illustration of the construction of the auxiliary graph $H$.

\vspace{-4mm}

We use a standard matching algorithm to find a maximum matching $M_H$ between $V_1$ and $V_2$ in $H$.

\textbf{Case 2.1}: All vertices in $V_1$ are matched in $M_H$. For this case, we will show that we can find a group-satisfied allocation to allocate the whole graph $G$ according to the matching $M_H$.
First of all, the vertex $p_r\in V_1$ can only be matched with $i_0\in V_2$ since $p_r$ is only adjacent to $i_0$ in $H$. So we assign the bundle $p_r$ to agent $i_0$.
Next, we consider other edges in the matching $M_H$.
Assume that $c_{j}\in C$ is matched with $i_j \in V_2$. Then agent $i_j$ is a dominator of the subtree $c_j$. We consider an $x_j$-partition of $c_j$ by agent $i_j$ and assign the bundle containing the center vertex to agent $i_j$. All other bundles left are bundles of a single chore.
For each vertex in  $c_{j}\in C$, we do the above allocation. Then we can allocate $|M_H|=|C|+1$ bundles to $|M_H|$ different agents since $M_H$ is a matching.
After this, the remaining chores will form an independent set. We assign each remaining chore to a remaining agent. This is the algorithm.

\begin{lemma}\label{gs_case2.1}
The allocation in Case 2.1 is group-satisfied.
\end{lemma}
\begin{proof}
    First, we show that the allocation is a valid allocation to allocate all chores to agents. The MMS$_{i_0}$ split $P_{i_0}$ of agent $i_0$ splits the graph into $n$ bundles.
    For a star $c_{i}\in C$, which is  $x_i$-super,  if it contains exact $x'$ bundles in $P_{i_0}$, then according to Definition~3 we know that $x_i\leq x'$. This relation holds for all stars in $C$.
    In our allocation, we will split each star $c_{i}\in C$ into $x_i\leq x'$ bundles. So our allocation will split the whole graph into $n'\leq n$ bundles. In our allocation, the first $|C|+1$ bundles are assigned according to
    the matching. So no two bundles are assigned to the same agent. For the remaining bundles, each of them contains a single chore and the number of bundles is not greater than the number of remaining agents by $n'\leq n$.
    So all chores can be assigned to agents. This is a valid allocation to allocate all chores to agents.

    Second, we show that all agents are satisfied with the allocation. First, agent $i_0$ is satisfied with the bundle $p_r$ because $p_r$ is a bundle in his MMS$_{i_0}$ split $P_{i_0}$.
    For each $c_{i}\in C$, we will assign a bundle containing the center vertex of $c_i$ to agent $i_j$. Agent $i_j$ is satisfied with this bundle because the star $c_i$ was split by agent $i_j$.
    All other bundles left are bundles of a single chore. Each agent is satisfied with a single chore. So all agents are satisfied with the allocation.
\end{proof}

\textbf{Case 2.2}: Some vertices in $V_1$ are not in the matching $M_H$. In this case, we will show that we can find a group-satisfied allocation to allocate a subset of stars at $C$.
Our algorithm will use the following concept \emph{crown}.

\begin{definition}
Let $H=(V_1, V_2, E_H)$ be a bipartite graph.
A pair of nonempty vertex sets $(V'_1,V'_2)$ is called a \emph{crown} of $H$, if the following conditions hold:
\begin{enumerate}
\item $V'_1\subseteq V_1$ and $V'_2 \subseteq V_2$.
\item any vertex in $V'_1$ is only adjacent to vertices in $V'_2$.
\item there is a matching $M'_H$ of size $|V'_2|$ between $V'_1$ and $V'_2$. The matching  $M'_H$ is called a \emph{witness matching} of the crown.
\end{enumerate}
\end{definition}

The following lemma gives a condition for the existence of the crown structure.

\begin{lemma}\label{condition_crown}
Let $H=(V_1, V_2, E_H)$ be a bipartite graph and $M_H$ be a maximum matching between $V_1$ and $V_2$. If $|M_H|< |V_1|$, then there is a crown $(V'_1,V'_2)$ of $H$, which can be found in linear time.
\end{lemma}
\begin{proof}
    A vertex in $H$ is called \emph{$M_H$-saturated} if it is an endpoint of an edge in $M_H$ and \emph{$M_H$-unsaturated} otherwise. A path $P$ in $H$ that alternates between edges not in $M_H$ and edges in $M_H$ is called an \emph{$M_H$-alternating path}.
    Since  $|M_H|< |V_1|$, we know that there are some $M_H$-unsaturated vertices in $V_1$.
    Let $V'_1\subseteq V_1$ be the set of vertices in $V_1$ that are reachable from an $M_H$-unsaturated vertex in $V_1$ via an $M_H$-alternating path, possibly of length zero
    (which means that all $M_H$-unsaturated vertices in $V_1$ are in $V'_1$).
    Let $V'_2\subseteq V_2$ be the set of vertices in $V_2$ that are reachable from an $M_H$-unsaturated vertex in $V_1$ via an $M_H$-alternating path.
    Note that $V'_1$ and $V'_2$ can be computed in linear time by a breadth-first search if $M_H$ is given.
    We will show that $(V'_1,V'_2)$ is a crown.

    The first condition in the definition of crown is trivial. The second condition holds because any vertex $v$ adjacent to a vertex $u\in V'_1$ must be in $V'_2$ because any $M_H$-alternating path from an $M_H$-unsaturated vertex in $V_1$ to $u$ plus the edge $uv$ will form an $M_H$-alternating path from an $M_H$-unsaturated vertex in $V_1$ to $v$, no matter $uv$ is in $M_H$ or not.
    For the third condition, we show that the subset $M'_H$ of edges in $M_H$ with two endpoints in $V'_1\cup V'_2$ will form a witness matching.
    We know that $M'_H$ is a matching since it is a subset of the matching $M_H$. We only need to prove that $|M'_H|=|V'_2|$. It is sufficient to prove that any vertex in $V'_2$ is contained in an edge in $M'_H$.
     For any $v' \in V'_2$,  there is an $M_H$-alternating path $P$ from an $M_H$-unsaturated vertex $u'\in V_1$ to $v'$. Note that the first edge (containing $u'$) in $P$ is not in $M_H$ since $u'$ is  $M_H$-unsaturated.
     If the last edge (containing $v'$) in $P$ is not in $M_H$, then we could get a bigger matching $M^*_H$ of $H$ by replacing $M_H\cap E(P)$ with $E(P)\setminus M_H$ in $M_H$,
     which is a contradiction to the maximality of $M_H$. So we know that any vertex in $V'_2$ is contained in an edge in $M'_H$.
     All three conditions in the definition of the crown hold. So $(V'_1,V'_2)$ is a crown.
    \end{proof}

\begin{lemma} \label{crownredu}
Let $H=(V_1, V_2, E_H)$ be the auxiliary bipartite graph constructed in Case~2 for trees with depth 3.
Given a crown $(V'_1,V'_2)$ of $H$, we can find a  group-satisfied allocation in linear time.
\end{lemma}

\begin{proof}
    We will use Lemma 3
    to prove this lemma. So we only need to show that we can always find a group-satisfied allocation after executing which the remaining graph is still a connected tree or an empty graph.

    If the condition of Step 2 holds, then it becomes Case~1. For this case, we find an $x$-perfect subtree.
    We will assign the $x$-perfect subtree to $x$ agents such that all the $x$ agents are satisfied with this allocation.
    By Lemma 4, we know that this allocation is group-satisfied.
    Otherwise, the algorithm will execute Step 5, and then it becomes Case~2. If the condition of Step 7 holds (Case 2.1), we will get a group-satisfied allocation by Lemma~\ref{gs_case2.1}.
    If the condition of Step 9 holds (Case 2.2), we will also get a group-satisfied allocation by Lemma~\ref{crownredu}. In any case, we can get a group-satisfied allocation that keeps the remaining graph connected.
    By Lemma 3, we know that this lemma holds.
    \end{proof}

For Case 2.2, the size of the matching $M_H$ is less than the size of $V_1$, and then the condition of Lemma~\ref{condition_crown} holds.
By Lemma~\ref{condition_crown}, we can find a crown $(V'_1,V'_2)$ in polynomial time.
Then we execute the group-satisfied allocation according to the crown $(V'_1,V'_2)$ by Lemma~\ref{crownredu}.

The main steps of our algorithm to compute MMS allocations of chores on trees with depth 3 are described as Algorithm~\ref{alg:two}.
\begin{algorithm}[h!]
    \caption{Depth$(A,C,U,G)$}
    \label{alg:two}
    \KwIn{An instance $I=(A,C,U,G)$, where $G=(C,E)$ is a tree with depth 3.}
    \KwOut{An MMS allocation of $I$.}
    Select an arbitrary agent $i_0$ and let the MMS$_{i_0}$ split be $P_{i_0}$.
    Let $p_r \in P_{i_0}$ be the bundle containing the root $r$.
    Let $C=\{c_1,c_2,\dots,c_p\}$ be the set of star components in $G'=G-p_r$\;
    \If{A star $c_q\in C$ is $x$-perfect for some integer $x$,}
    {Assign the star $c_q$ to $x$ agents: split $c_q$ into $x$ bundles according to an agent $i$ who $x$-partitions it, and assign
    the bundle containing the center vertex of $c_q$ to agent $i$ and assign the left $x-1$ chores to arbitrary $x-1$ agents\; and \textbf{return} Depth$(I')$, where $I'$ is the remaining instance after assigning $c_p$ to $x$ agents\;}
    \Else{Each star in $C$ is $x$-super for some integer $x$; We compute the auxiliary graph $H=(V_1, V_2, E_H)$ and a maximum matching $M_H$ in $H$\;
    \If{ $|V_1|=|M_H|$}{Assign the whole graph $G$ according to $M_H$\;}
    \Else{Compute the crown structure in $H$ and assign some stars in $C$ to agents according to the crown\; and \textbf{return} Depth$(I')$, where $I'$ is the remaining instance after the assignment.}
    }
\end{algorithm}

\begin{lemma}\label{depthalg}
MMS allocations of chores in trees with depth 3 always exist and can be computed in polynomial time.
\end{lemma}

\begin{proof}
    We will use Lemma~\ref{important} to prove this lemma. So we only need to show that we can always find a group-satisfied allocation after executing which the remaining graph is still a connected tree or an empty graph.

   If the condition of Step 2 holds, then it becomes Case~1. For this case, we find an $x$-perfect subtree.
    We will assign the $x$-perfect subtree to $x$ agents such that all the $x$ agents are satisfied with this allocation. By Lemma~\ref{perfect_c}, we know that this allocation is group-satisfied.
    Otherwise, the algorithm will execute Step 5, and then it becomes Case~2. If the condition of Step 7 holds (Case 2.1), we will get a group-satisfied allocation by Lemma~\ref{gs_case2.1}.
    If the condition of Step 9 holds (Case 2.2), we will also get a group-satisfied allocation by Lemma~\ref{crownredu}. In any case, we can get a group-satisfied allocation that keeps the remaining graph connected.
    By Lemma~\ref{important}, we know that this lemma holds.
    \end{proof}

\section{Spiders}
A graph is a \emph{spider} if it is a tree having only one vertex of degree $\geq 3$.
We will also use the group-satisfied method to show the existence of MMS allocations of chores in spiders.

The vertex of degree $\geq 3$ in a spider is called the \emph{center}. A degree-1 vertex in a spider is called a \emph{leaf}.
The path from a leaf to the center is called
a \emph{branch} of the spider and the number of edges in a branch is the \emph{length} of the branch.

In this section, we assume that the input graph $G$ is a spider,
and use $r$ to denote the center and use $B_i$ to denote the branch between the center and a leaf $f_i$.
We also consider the tree as a rooted tree with the root being the center $r$.

Similar to the algorithm for trees with depth 3, the algorithm will also use Lemma~\ref{perfect_c} and Lemma~\ref{super_c} to construct group-satisfied partial allocations based on $x$-perfect and $x$-super subtrees.

\begin{lemma}\label{spiderbranch}
If there is a branch $B_i$ of $G$ not completely contained in one bundle in the MMS$_j$ split of any agent $j\in A$, then we can find a group-satisfied allocation after which the remaining graph is still a spider.
\end{lemma}
    \begin{proof}
    We give an algorithm to prove this lemma.
    For each agent $j$, the bundle in the MMS$_j$ split containing a leaf is called an \emph{ending bundle}. For each agent, the ending bundle containing the leaf $f_i$ is a subpath of the branch $B_i$ by the condition of the lemma. Assume that agent $j_0$ has the longest ending bundle containing $f_i$. We assign this bundle to the agent $j_0$.
    This allocation is group-satisfied because all agents are satisfied with this allocation and the remaining graph is still a spider. This algorithm is like what we do in paths.
    \end{proof}

The above lemma provides a way to find group-satisfied allocations for a special case. Our algorithm iteratively applies the operation in the proof of Lemma~\ref{spiderbranch} until we get a spider such that each branch of it is contained in one bundle in the MMS$_j$ split of an agent $j$. The following part of the algorithm is similar to the algorithm for
trees with depth 3.

We consider the MMS$_{j_0}$ split $P_{j_0}$ of an arbitrary agent $j_0$ and denote the bundle containing the root by $p_r$.
In graph $G'=G-p_r$, each connected component is a path.
Let $C=\{c_1,c_2,\dots,c_p\}$ be the set of these paths. Each path in $C$ is 1-super because the whole branch is contained in a bundle in the MMS$_{j}$ split of some agent $j$.

We also construct an auxiliary bipartite graph $H=(V_1, V_2, E_H)$.
Set $V_1$ contains $|C|+1$ vertices, which are corresponding to the $|C|$ paths in $C$ and the bundle $p_r$.
Each vertex in $V_2$ is corresponding to an agent in $A$ and we simply denote $V_2$ by $A$. A vertex $j\in V_2$ is adjacent to a vertex $c\in C$ in $H$ if and only if the corresponding agent $j\in A$ is a dominator of $c$.
Vertex $p_r\in V_1$ is only adjacent to $j_0\in A$. We compute a maximum matching $M_H$ between $V_1$ and $V_2$ in $H$.

\textbf{Case 1}: $|M_H|=|V_1|$. We allocate the whole graph to the $n$ agents according to the matching $M_H$. The bundle $c$ corresponding to a vertex in $V_1$ will be assigned to agent $i$ if they are matched in $M_H$.
It is easy to see that all chores will be allocated to agents, and all agents are satisfied with the allocation. The allocation is group-satisfied.

\textbf{Case 2}: $|M_H|<|V_1|$.
By Lemma~\ref{condition_crown}, we can find a crown $(V'_1,V'_2)$ in polynomial time.
We allocate the chores to agents according to the algorithm in the proof of the following lemma.

\begin{lemma}  \label{crownredu-s}
Let $H=(V_1, V_2, E_H)$ be the auxiliary bipartite graph constructed in Case~2 for spiders.
Given a crown $(V'_1,V'_2)$ of $H$, we can find a group-satisfied allocation in linear time.
\end{lemma}
 \begin{proof}
        We will give an algorithm to find the group-satisfied allocation. The proof is similar to the proof of Lemma~8.

        Let $M'_H$ be the witness matching of the crown $(V'_1,V'_2)$. Let $V^*_1\subseteq V'_1$ be the subset of vertices appearing in $M'_H$. The
        group-satisfied allocation will assign each path in  $V^*_1$ to an agent in $V'_2$.
        We assume that $c_{j}\in V^*_1$ is matched with $i_j \in V'_2$ in $M'_H$.
        Then agent $i_j$ is a dominator of the subpath $c_j$ and agent $i_j$ is satisfied with $c_j$. We assign $c_j$ to agent $i_j$.
        Then we allocate each subpath in $V^*_1$ to a different agent in $V'_2$ since $M'_H$ is a matching. Furthermore, all agents in $V'_2$ are assigned with a bundle.
        This is the allocation algorithm. It is easy to see that the algorithm can be executed in linear time.

        We show that the allocation satisfies the condition in Lemma~5 to prove that it is group-satisfied.
        It is easy to see that after deleting all the subpaths in $V^*_1$ from $G$, the remaining graph is still a spider.
        According to the construction of $H$ and the definition of crown, we know that for any path $c_{j}\in V^*_1$, the set of dominators of it is a subset of $V'_2$, i.e., $D(c_j)\subseteq V^*_2$ holds for each $c_{j}\in V^*_1$.
        In our algorithm, all agents in $V'_2$ will be assigned with a bundle. So any agent left unassigned is not a dominator of a path in $V^*_1$. By Lemma~5, we know that the allocation is group-satisfied.
        \end{proof}

All the above results imply that

\begin{lemma} \label{spideralg}
MMS allocations of chores on spiders always exist and can be computed in polynomial time.
\end{lemma}
\begin{proof}
    Our algorithm to find the ${\frac{3}{2}}$-MMS allocation is that: first, remove an arbitrary edge from the cycle, and then find an MMS allocation for the instance on a path.
    To prove the correctness, we only need to show that the for each agent the MMS value on the path is not less than ${\frac{3}{2}}$ times of his MMS value on the cycle, since we can always find MMS allocations for chores on paths in polynomial time by using the algorithm presented in previous sections.

    For each agent $i$, we consider the MMS$_i$ split $P_i$ of the cycle $G$. After deleting an edge $e$ from $G$, one bundle in $P_i$ may be split into two pieces $x_1$ and $x_2$. At least one piece, say $x_1$ has disutility not less than ${\frac{1}{2}}mms_i(G)$. We adjoin $x_1$ with its neighbor bundle in $P_i$ and let $x_2$ become a single bundle. In this way, we split the path $G-e$ into $n$ bundles, the disutility of each bundle is not less than ${\frac{3}{2}} mms_i(G)$. Then we know that the $mms_i(G-e)\geq {\frac{3}{2}} mms_i(G)$.
\end{proof}


\section{Cycles}
Next, we consider the case where the chores are on a simple cycle. Different from trees, MMS allocations of chores on cycles may not exist.
We will give an example to show the nonexistence of MMS allocations even to only three agents. We will show how we find the example later.

\begin{table}[h]
    \centering
    \begin{tabular}{p{40pt}|p{12pt}|p{12pt}|p{12pt}|p{12pt}|p{12pt}|p{12pt}|p{12pt}|p{12pt}|p{12pt}}
        \hline
         &$c_1$&$c_2$&$c_3$&$c_4$&$c_5$&$c_6$&$c_7$&$c_8$&$c_9$\\
        \hline
        Agent 1&$-\frac{1}{2}$&$-\frac{1}{3}$&$-\frac{1}{6}$&$-\frac{1}{6}$&$-\frac{1}{3}$&$-\frac{1}{2}$&$-\frac{1}{3}$&$-\frac{1}{3}$&$-\frac{1}{3}$\\
        \hline
        Agent 2&$-\frac{1}{6}$&$-\frac{1}{2}$&$~0$&$-\frac{1}{2}$&$-\frac{1}{6}$&$-\frac{1}{2}$&$-\frac{1}{3}$&$-\frac{1}{3}$&$-\frac{1}{2}$\\
        \hline
        Agent 3&$-\frac{1}{3}$&$-\frac{1}{6}$&$-\frac{1}{6}$&$-\frac{1}{3}$&$-\frac{1}{2}$&$-\frac{1}{3}$&$-\frac{1}{3}$&$-\frac{1}{3}$&$-\frac{1}{2}$\\
        \hline
    \end{tabular}
    \caption{An example of nonexistence of $\alpha$-MMS allocations of a 9-cycle to three agents for any $\alpha <{\frac{7}{6}}$}\label{t1-ex}
\end{table}

In the example in Table~\ref{t1-ex}, we are going to allocate nine chores on a cycle to three agents. The chores appear on the cycle are in the same order as that in the table. The numbers in the table are the disutilities of the chores for the agents. The MMS value of each agent is $-1$. For any four consecutive chores, the disutility for any agent is at most  $-\frac{7}{6}$. If an $\alpha$-MMS allocation with $\alpha <\frac{7}{6}$ exists, then it would split the cycle into
three bundles of three consecutive chores. There are only three ways to split it. It is easy to check that none of the three partitions yields an $\alpha$-MMS allocation with $\alpha <\frac{7}{6}$.
Thus, no $\alpha$-MMS allocation with $\alpha <\frac{7}{6}$ exists for this instance.

On the other hand, ${\frac{3}{2}}$-MMS allocations of chores on cycles always exist.
The simple observation of the ${\frac{1}{2}}$-MMS allocations of goods on cycles in~\cite{LT18} can get the corresponding result for chores.

\begin{lemma}
An ${\frac{3}{2}}$-MMS allocation of chores on a cycle always exists and can be found in polynomial time.
\end{lemma}
 \begin{proof}
    Our algorithm to find the ${\frac{3}{2}}$-MMS allocation is that: first, remove an arbitrary edge from the cycle, and then find an MMS allocation for the instance on a path.
    To prove the correctness, we only need to show that the for each agent the MMS value on the path is not less than ${\frac{3}{2}}$ times of his MMS value on the cycle, since we can always find MMS allocations for chores on paths in polynomial time by using the algorithm presented in previous sections.

    For each agent $i$, we consider the MMS$_i$ split $P_i$ of the cycle $G$. After deleting an edge $e$ from $G$, one bundle in $P_i$ may be split into two pieces $x_1$ and $x_2$. At least one piece, say $x_1$ has disutility not less than ${\frac{1}{2}}mms_i(G)$. We adjoin $x_1$ with its neighbor bundle in $P_i$ and let $x_2$ become a single bundle. In this way, we split the path $G-e$ into $n$ bundles, the disutility of each bundle is not less than ${\frac{3}{2}} mms_i(G)$. Then we know that the $mms_i(G-e)\geq {\frac{3}{2}} mms_i(G)$.
\end{proof}


\subsection{Allocating a cycle to three agents}\label{cyclechores}
We use a linear programming method to help us find the (tight) approximation ratio for allocating chores on a cycle to three agents.
We will construct a linear programming model for our problem. Let $\alpha$ be the ratio. Our objective is to find the maximum value of $\alpha$ such that
for any valid allocation of the cycle to $n$ agents, at least one agent will get a bundle with the disutility less than  $\alpha$ times of his MMS value.
So our objective is to find the maximum value of $\alpha$ such that there is no $\alpha$-MMS allocation.
However, the constraints in our model are hard to list out directly. The number of constraints may even be exponential to $n$ and $m$. We will first give some properties that will allow us to dramatically decrease the number of constraints if there are only three agents.
In the remaining of this subsection, we always assume that the problem is to allocate chores on a cycle to three agents.

\begin{lemma}\label{subsetcc}
Assume that the instance has three agents and let $\alpha\geq1$. If there is an $\alpha$-MMS$_i$ split $P_i$ of $G$ for agent $i$ and an $\alpha$-MMS$_j$ split $P_j$ of $G$ for agent $j$ such that one bundle in $P_i$ is a subset of one bundle in $P_j$,
then  $\alpha$-MMS allocations for this instance always exist.
\end{lemma}
\begin{proof}
    Let the agent set be $\{1,2,3\}$ and $P_i=\{B_{i1},B_{i2},B_{i3}\}$ be an $\alpha$-MMS$_i$ split for agent $i$ ($i\in\{1,2,3\}$).
    W.l.o.g, we assume that the bundle $B_{21}$ in $P_2$ is a subset of the bundle $B_{11}$ in $P_1$ and show
     that there is an $\alpha$-MMS allocation. We partition the cycle into three connected bundles $B_{11}$, $B'_{22}=B_{22}-B_{11}$, and $B'_{23}=B_{23}-B_{11}$. We will assign the three bundles to different agents according to different cases.

    Case 1. $u_3(B_{11})\leq mms_3$: Now we have that $u_3(G-B_{11})\geq 2 mms_3$. For any partition of $G-B_{11}$ into two connected bundles, the disutility of at least one bundle is not less than $mm_3$ for agent 3. So
    one of $u_3(B'_{22}) \geq  mms_3$ and $u_3(B'_{23}) \geq  mms_3$ always holds.
    Note that $B'_{22}$ and $B'_{23}$ are subsets of $B_{22}$ and $B_{23}$. We know that $u_2(B'_{22}) \geq \alpha \cdot mms_2$ and $u_2(B'_{23}) \geq \alpha \cdot mms_2$. We assign $B_{11}$ to agent 1, let agent 3 takes one of the favorite in  $B'_{22}$
    and $B'_{23}$, and assign the remaining bundle to agent 2. This is an $\alpha$-MMS allocation.

    Case 2. $u_3(B_{11})> mms_3$: We can see that either $B_{12}\supseteq B'_{22}$ or $B_{13}\supseteq B'_{23}$ holds since
    $B_{12}\cup B_{13} =B'_{22}\cup B'_{23}$. So one of $u_1(B'_{22}) \geq \alpha \cdot mms_1$ and $u_1(B'_{23}) \geq \alpha \cdot mms_1$ always holds. Then we assign $B_{11}$ to agent 3, let agent 1 takes one of the favorite in  $B'_{22}$
    and $B'_{23}$, and assign the remaining bundle to agent 2. This is an $\alpha$-MMS allocation.
\end{proof}

A partition of the cycle $G$ into 3 bundles can be represented by a set of three edges in the cycle.
We will call it the \emph{cut representation} of the partition.
Next, we give some important properties for instance having no $\alpha$-MMS allocation, which will be used to build our linear programming constraints.

\begin{figure}[htbp]
    \centering
    \begin{tikzpicture}
        [scale = 1, line width = 0.5pt,solid/.style = {circle, draw, fill = black, minimum size = 0.3cm},empty/.style = {circle, draw, fill = white, minimum size = 0.3cm}]
        \node[empty,label=center:\tiny$S_1$] (B1) at (90:1cm) {};
        \node[empty,label=center:\tiny$S_2$] (C1) at (50:1cm) {};
        \node[empty,label=center:\tiny$S_3$] (D1) at (10:1cm) {};
        \node[empty,label=center:\tiny$S_4$] (E1) at (330:1cm) {};
        \node[empty,label=center:\tiny$S_5$] (F1) at (290:1cm) {};
        \node[empty,label=center:\tiny$S_6$] (G1) at (250:1cm) {};
        \node[empty,label=center:\tiny$S_7$] (H1) at (210:1cm) {};
        \node[empty,label=center:\tiny$S_8$] (I1) at (170:1cm) {};
        \node[empty,label=center:\tiny$S_9$] (J1) at (130:1cm) {};

        \node[label=center:\tiny$e_{11}$] (B2) at (110:1.2cm) {};
        \node[label=center:\tiny$e_{21}$] (B2) at (70:1.2cm) {};
        \node[label=center:\tiny$e_{31}$] (B2) at (30:1.2cm) {};
        \node[label=center:\tiny$e_{12}$] (B2) at (350:1.2cm) {};
        \node[label=center:\tiny$e_{22}$] (B2) at (310:1.2cm) {};
        \node[label=center:\tiny$e_{32}$] (B2) at (270:1.2cm) {};
        \node[label=center:\tiny$e_{13}$] (B2) at (230:1.2cm) {};
        \node[label=center:\tiny$e_{23}$] (B2) at (190:1.2cm) {};
        \node[label=center:\tiny$e_{33}$] (B2) at (150:1.2cm) {};





        \draw[red] (B1)--(C1);
        \draw[red] (E1)--(F1);
        \draw[red] (H1)--(I1);

        \draw[blue] (C1)--(D1);
        \draw[blue] (F1)--(G1);
        \draw[blue] (I1)--(J1);

        \draw[black,dashed] (J1)--(B1);
        \draw[black,dashed] (D1)--(E1);
        \draw[black,dashed] (G1)--(H1);

    \end{tikzpicture}
    \caption{A cycle, where the set of three black edges $P_1=\{e_{11}, e_{12}, e_{13}\}$ is the cut representation of an MMS$_1$ split for agent $1$, the set of three red edges $P_2=\{e_{21}, e_{22}, e_{23}\}$ is the cut representation of an MMS$_2$ split for agent $2$, the set of three blue edges $P_3=\{e_{31}, e_{32}, e_{33}\}$ is the cut representation of an MMS$_3$ split for agent $3$, and $\{S_1,S_2,\dots, S_9\}$ are the nine segments after deleting $P_1\cup P_2 \cup P_3$. }
    \label{9-cycle}
\end{figure}
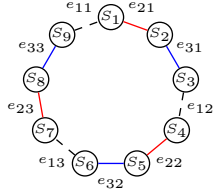

\begin{lemma} \label{lpuse1}
Assume that the instance has three agents. Let $P_1=\{e_{11}, e_{12}, e_{13}\}, P_2=\{e_{21}, e_{22}, e_{23}\}$ and $P_3=\{e_{31}, e_{32}, e_{33}\}$ be the cut representations of $\alpha$-MMS splits of $G$ for agents 1, 2 and 3, respectively, which always exist for $\alpha \geq 1$.
If the instance has no $\alpha$-MMS allocation, then it holds that\\
(a) all the edges in $P_1, P_2$ and $P_3$ are different;\\
(b) we can relabel the edges in the three sets $P_1, P_2$ and $P_3$ such that the nine edges appear in the following order
on the cycle $G$: $e_{11}e_{21}e_{31}e_{12}e_{22}e_{32}e_{13}e_{23}e_{33}$ (See Figure~\ref{9-cycle}).
\end{lemma}

\begin{proof}
    (a) Assume that two of $P_1$, $P_2$ and $P_3$, say $P_1$ and $P_2$ have a common edge $e$.
    After deleting $e$ the graph becomes a path. Both of $P_1$ and $P_2$ will partition the path into three connected bundles.
    Let $b_1$ be the bundle of agent 1 containing the most left chore on the path $G-e$ and let $b_2$ be the bundle of agent 2 containing the most left chore on $G-e$. It is easy to see that either $b_1\subseteq b_2$ or $b_2\subseteq b_1$,
    which implies the condition of Lemma~\ref{subsetcc} holds. We get a contradiction that the instance has an $\alpha$-MMS allocation.
    So the edges in $P_1, P_2$ and $P_3$ are different.

    (b) If the nine edges do not appear in the above order, say exchanging the positions of any two edges, we will see that one bundle of one agent is a subset of one bundle of another agent. By Lemma~\ref{subsetcc}, we will also get a contradiction.
    We omit the trivial and tedious case analysis here.
    \end{proof}

\begin{corollary}
MMS allocations of at most eight chores on a cycle to three agents always exist.
\end{corollary}
\begin{proof}
    When there are at most eight chores, the cycle $G$ has at most eight edges.  The cut representations of MMS split of $G$ for agents 1, 2, and 3 will have at least one common edge. By Lemma~15(a),
    we know that MMS allocation always exists.
\end{proof}

Let $P_1$, $P_2$ and $P_3$ be the edge representations of an MMS split of $G$ for agents 1, 2 and 3, respectively.
Each connected part in the graph after deleting edges in $P_1\cup P_2 \cup P_3$ is called a \emph{segment} (See Figure~\ref{9-cycle}).

\begin{lemma}\label{lpuse1-1}
Assume that the instance has three agents. Fix an MMS$_i$ split of $G$ for each agent $i$. Let $\alpha\geq 1$.
If the instance has no $\alpha$-MMS allocation, the disutility of any four consecutive segments is less than $\alpha\cdot mms_i$ for any agent $i$.

\end{lemma}

\begin{proof}
    Let $b$ denote four consecutive segments. Assume that $u_{i_0}(b)\geq \alpha \cdot mms_{i_0}$ holds for an agent $i_0\in N$.
    We show that there is always an $\alpha$-MMS allocation.

    Case 1. $b$ contains a bundle in the MMS$_{i_0}$ split of $G$ for agent $i_0$: Then there is an $\alpha$-MMS$_{i_0}$ split for agent $i_0$ such that one bundle in it is exactly $b$.
    Note that any four consecutive segments will contain two bundles in the MMS splits of $G$ for two different agents. Any MMS split is also an $\alpha$-MMS split for $\alpha \geq 1$.
    So the condition of Lemma~\ref{subsetcc} holds and then $\alpha$-MMS allocations always exist.

    Case 2. $b$ does not contain a bundle in the MMS$_{i_0}$ split of $G$ for agent $i_0$. For this case, it is easy to see that the remaining path $G-b$ consists of two connected paths, each of which is a subset of a bundle
    in the MMS split of $G$ for an agent other than $i_0$. We assign these two parts to these two agents and assign $b$ to agent $i_0$. This will be an $\alpha$-MMS allocation.

    In any case, we can find an $\alpha$-MMS allocation, a contradiction. So the lemma holds.
    \end{proof}

\begin{lemma}\label{lpuse2}
Assume that the instance has three agents. Let $\alpha\geq 1$.
 If the instance has no $\alpha$-MMS allocation, then for any $\alpha$-MMS$_i$ split $P_i$ of $G$ for agent $i\in A$, there is one bundle $b$ in $P_i$ such that $u_j(b)\geq \alpha \cdot mms_j$ holds for any agent $j\in A\setminus \{i\}$, and for all other bundles $b'$ in $P_i$ it holds that $u_j(b')< \alpha \cdot mms_j$ for any agent $j\in A\setminus \{i\}$.
\end{lemma}

\begin{proof}
    We say that an agent is satisfied with a bundle if the disutility of this bundle is not less than $\alpha$ times of his MMS value.
    We know that agent $i$ is satisfied with all the three bundles in $P_i$.
    For the other two agents $\{j_1,j_2\}=A\setminus \{i\}$, each is satisfied with at least one bundle in $P_i$.
    If agent $j_1$ and agent $j_2$ are satisfied with two different bundles in $P_i$, then we assign these two bundles in $P_i$ to them and assign the last bundle in $P_i$ to agent $i$.
    This would be an $\alpha$-MMS allocation. So we know that agent $j_1$ and agent $j_2$ are satisfied with the same bundle in $P_i$.
\end{proof}

Now, we are ready to describe our linear programming (LP). Assume that the instance has no $\alpha$-MMS allocation. We use Lemmas~\ref{lpuse1} to~\ref{lpuse2} to construct the constraints in LP.

We consider an instance $I=(A,C,U,G)$ of allocating chores on a cycle $G$ to three agents $A=\{1,2,3\}$.
Let $P_1=\{e_{11}, e_{12}, e_{13}\}, P_2=\{e_{21}, e_{22}, e_{23}\}$ and $P_3=\{e_{31}, e_{32}, e_{33}\}$ be the cut representations of MMS splits of $G$ for agents 1, 2 and 3, respectively.
We assume that $I$ does not have $\alpha$-MMS allocation for some constant $\alpha> 1$. By Lemma~\ref{lpuse1}, we know that the nine edges in $P=P_1\cup P_2 \cup P_3$ are different and
we can assume without loss of generality that they appear in the following order
on the cycle $G$: $e_{11}e_{21}e_{31}e_{12}e_{22}e_{32}e_{13}e_{23}e_{33}$.
After deleting the nine edges in $P$, the cycle $G$ will be split into nine segments. We label the nine segments as $S_1, S_2,\dots, S_9$ in the order as shown in Figure~\ref{9-cycle}.
Let the MMS splits of $G$ for agents 1, 2 and 3 be $P_1=\{B_{11}, B_{12}, B_{13}\}, P_2=\{B_{21}, B_{22}, B_{23}\}$ and $P_3=\{B_{31}, B_{32}, B_{33}\}$, where $B_{11}=S_1\cup S_2 \cup S_3$, $B_{12}=S_4\cup S_5 \cup S_6$,
$B_{13}=S_7\cup S_8 \cup S_9$, $B_{21}=S_2\cup S_3 \cup S_4$, $B_{22}=S_5\cup S_6 \cup S_7$, $B_{23}=S_8\cup S_9 \cup S_1$, $B_{31}=S_3\cup S_4 \cup S_5$, $B_{32}=S_6\cup S_7 \cup S_8$,
and $B_{33}=S_9\cup S_1 \cup S_2$.

For each segment, it may contain only one chore. So in the next analysis, we will not split a segment anymore and consider each segment as a single big chore.
Besides $\alpha$, our LP has $3\times9$ variables $x_j, y_j$ and $z_j$ ($j\in \{1,2,\dots, 9\}$). The variable $x_j$ (resp., $y_j$ and $z_j$) is the disutility of segment $S_j$ of agent 1 (resp., agent 2 and agent 3), i.e.,
$x_j=u_1(S_j)$, $y_j=u_2(S_j)$ and $z_j=u_3(S_j)$. The first set of constraints in our LP model is to set the domain of these variables.
\begin{eqnarray}\label{equs1}
    x_j,y_j,z_j\leq 0 ~~~~~~(j=1,2,\dots,9).
\end{eqnarray}

We normalize the MMS value for each agent by letting $mms_1=mms_2=mms_3=-1$. According to the MMS splits of the three agents, we get $3\times 3$ constraints
\begin{eqnarray}
    x_{1+i}+x_{2+i}+x_{3+i}\geq -1&(i=0,3,6);\label{equs2}\\
    y_{2+i}+y_{3+i}+y_{4+i}\geq -1&(i=0,3,6);\label{equs3}\\
    z_{3+i}+z_{4+i}+z_{5+i}\geq -1&(i=0,3,6),\label{equs4}
\end{eqnarray}
where the indices are computed modulo 9. We will always assume this without stating it every time.

By Lemma~\ref{lpuse1-1}, we directly get $3\times 9$ constraints
\begin{eqnarray}
    x_{1+i}+x_{2+i}+x_{3+i}+x_{4+i}< -\alpha&(i=0,1,\dots,8);\label{equs5}\\
    y_{1+i}+y_{2+i}+y_{3+i}+y_{4+i}< -\alpha&(i=0,1,\dots,8);\label{equs6}\\
    z_{1+i}+z_{2+i}+z_{3+i}+z_{4+i}< -\alpha&(i=0,1,\dots,8).\label{equs7}
\end{eqnarray}

Next, we consider the constraints generated by Lemma~\ref{lpuse2}. The three bundles in the MMS$_1$ split $P_1$ for agent 1 are $B_{11}, B_{12}$ and $B_{13}$.
By Lemma~\ref{lpuse2}, we know that for agents 2 and 3, at most one bundle is satisfied (the disutility is not less than $\alpha$ times of his MMS value).
W.l.o.g., we assume that agents 2 and 3 are satisfied with $B_{11}$, which means that agents 2 and 3 are not satisfied with the other two bundles. Then we get 4 constraints
\begin{eqnarray}
    u_i(B_{12})< \alpha \times mms_i =-\alpha&(i=2,3);\label{eqd1}\\
    u_i(B_{13})< \alpha \times mms_i =-\alpha&(i=2,3). \label{eqd2}
\end{eqnarray}

We look at the MMS$_2$ split $P_2=\{B_{21}, B_{22}, B_{23}\}$ for agent 2 and the MMS$_3$ split $P_3=\{B_{31}, B_{32}, B_{33}\}$ for agent 3.
By Lemma~\ref{lpuse2} again, agent 1 and agent 3 are satisfied with one bundle in $P_2$, and agent 1 and agent 2 are satisfied with one bundle in $P_3$.
By identifying the symmetrical cases (Rotations of the three agents are also considered), we only need to consider three different cases.
Case 1: agent 1 and agent 3 are satisfied with $B_{21}$, and  agent 1 and agent 2 are satisfied with $B_{31}$;
Case 2: agent 1 and agent 3 are satisfied with $B_{21}$, and  agent 1 and agent 2 are satisfied with $B_{32}$  
Case 3: agent 1 and agent 3 are satisfied with $B_{23}$, and agent 1 and agent 2 are satisfied with $B_{32}$.
For each case, we generate $4\times 2$ constraints like (\ref{eqd1}) and (\ref{eqd2}). There are three cases. So we will get three LP models. Each LP model has the same number of constraints.

By solving the three LPs, we get that $\alpha <{\frac{8}{7}}$ for two LPs and $\alpha <{\frac{7}{6}}$ for the other one. The bigger one is ${\frac{7}{6}}$.
This means for any $\alpha \geq {\frac{7}{6}}$, our LP will have no solution. We get the following result.

\begin{lemma}
$\frac{7}{6}$-MMS allocations of chores on a cycle to 3 agents always exist and can be found in polynomial time.
\end{lemma}
\begin{proof}
    For any instance, we fix an MMS split $P_i$ for each agent $i$ (considering them as the cut representations). If they have some common edges, then we can find an MMS allocation directly in polynomial time by Lemma~\ref{lpuse1}(a) and the algorithm in Lemma~\ref{subsetcc}.
    Otherwise, we let $P_1=\{e_{11}, e_{12}, e_{13}\}, P_2=\{e_{21}, e_{22}, e_{23}\}$ and $P_3=\{e_{31}, e_{32}, e_{33}\}$. They are nine different edges.
    We can relabel them in the order as shown in Figure~4, otherwise, we know that one bundle of one agent will be a subset of one bundle of another agent by Lemma~\ref{lpuse1}(b) and we can find an MMS  allocation in polynomial time by Lemma~\ref{subsetcc}.

    Next, we can assume that the above LP model is suitable for this instance. The LP does not have any solution for $\alpha <{\frac{7}{6}}$, which means some constraints in the LP will not hold for $\alpha <{\frac{7}{6}}$.
    The constraints (1) to (4) clearly hold. If some constraints in (5) to (7) do not hold, then there is an $\frac{7}{6}$-MMS allocation with one bundle containing four consecutive segments by Lemma~\ref{lpuse1-1}.
    We can enumerate all these kinds of partitions and find the $\frac{7}{6}$-MMS allocation in polynomial time. Otherwise, one of (8) and (9) and the following constraints will not hold.
    For this case, one of $P_1$, $P_2$ and $P_3$ is a $\frac{7}{6}$-MMS allocation by Lemma~\ref{lpuse2} and we can also check it in polynomial time.

    In any case, there is a $\frac{7}{6}$-MMS allocation and it can be found in polynomial time.
\end{proof}

The LP model can also be used to find instances with possible tight approximation ratio. If we add the equal sign into the constraints (\ref{equs5}) to (\ref{eqd2}) and the following constraints, we will solve one LP with
$\alpha ={\frac{7}{6}}$ and two LPs with $\alpha ={\frac{8}{7}}$. For the LP with the solution $\alpha ={\frac{7}{6}}$, we get the values of the other variables $x_j, y_j$ and $z_j$ ($j\in \{1,2,\dots, 9\}$) as shown in Table~\ref{t1-ex}. That is the way we got the example of the nonexistence of $\alpha$-MMS allocations for $\alpha <{\frac{7}{6}}$ in Table~\ref{t1-ex}.

\begin{lemma}
For any $\alpha <{\frac{7}{6}}$, $\alpha$-MMS allocations of chores on a cycle to three agents may not exist.
\end{lemma}

The LP method can also be directly used to solve the problem of allocating goods on a cycle to three agents. The detailed arguments are put in Appendix~\ref{appen-2}. We get the following results which are consistent with the results obtained by using combinatorial analysis in~\cite{LT18}.
\begin{lemma}\label{lemgood}
$\alpha$-MMS allocations of goods on a cycle to three agents always exist and can be found in polynomial time for $\alpha \leq \frac{5}{6}$, and may not exist for $\alpha > \frac{5}{6}$.
\end{lemma}

\section{Discussion and Conclusion}
For MMS allocations of chores on a tree, we propose the group-satisfied method to solve the problem on two subclasses.
Whether MMS allocations of chores on general trees always exist or not is still an open problem.
We believe that MMS allocations of chores on trees always exist.
We also believe that the proposed group-satisfied method can be used to solve more related problems.

Another contribution of this paper is a novel method based on linear programming (LP) to characterize the optimal approximate MMS allocations without complicated combinatorial analysis. This method could potentially solve more general cases by figuring out simple and clean necessary conditions for the non-existence of $\alpha$-MMS allocation.

\bibliography{refs}
\bibliographystyle{abbrv}

\newpage

\appendix

\section{The proof of \Cref{MMScom}}\label{append-1}
In this section, we prove \Cref{MMScom} by giving algorithms to compute the MMS value for each agent for allocating chores on trees or cycles.
It is not hard to see that the computational task for cycles can be easily reduced to trees by enumerating
    one of the cutting edges and computing the corresponding MMS value on a path.

    Before introducing the algorithm for trees, we first give a subroutine to determine whether there is a valid $n$-partition $P^{n}$ of the chores $C$
    such that $u_i(P^{n}(j))\geq x$ for each $j\in [n]$. The details are listed in Algorithm~\ref{alg:check}. Recall that for a vertex $v$ in a rooted tree, we use $T_v$ to denote the subtree rooted
    at vertex $v$, use $\mathcal{C}_v$ to denote the children of vertex $v$ on a rooted tree, and denote $u_{i|C^{'}}$ as the disutility function restricted on the chores set $C^{'}$ for agent $i$.

    \begin{algorithm}[h]
        \caption{Check($G$, $u_i$, $n$, $x$)}
        \label{alg:check}
        \KwIn{An instance $I=(G, u_i, n ,x)$, where $G=(C,E)$ is a tree, $u_i$ is the disutility function of agent $i$,
         $n$ is the number of bundles and $x$ is the lower bound of any bundles.}
        \KwOut{To determine whether there is a valid $n$-partition such that the disutility of each bundle is no less than $x$.}
        \If{$u_i(C)\geq x$}{\textbf{return true}\;}
        \If{$n\leq 1$}{\textbf{return false}\;}
        Select an arbitrary vertex $r$ and consider the tree $G$ as a rooted tree with root $r$\;
        Find a vertex $c$ from top down to bottom such that $u_i(T_c)< x$ and $u_i(T_{c_j})\geq x$ for each child $c_j\in \mathcal{C}_{c} $ (we allow $\mathcal{C}_{c}=\emptyset$ )\;
        Let $c_{j_0}:=\arg \min_{c_j \in \mathcal{C}_{c}} u_i(T_{c_{j}})$ and $G^{'}=G\setminus T_{c_{j_0}}$\;
        \textbf{return} Check($G^{'},u_{i|G^{'}},n-1,x$)
    \end{algorithm}

    \begin{lemma}
        Given an instance $I=(G, u_i, n ,x)$, there exists a valid $n$-partition such that the disutility of each bundle is no less than $x$
        for agent $i$ if and only if Algorithm Check($G$, $u_i$, $n$, $x$) returns true.
    \end{lemma}

    \begin{proof}
        Firstly, it is easy to see that the decision algorithm presented above can be modified as a constructive algorithm. Thus, if the algorithm returns true,
        we can find an n-partition of the instance such that $u_i(P^{n}(j))\geq x$ for $j\in [n]$.

        It remains to show that for any instance $I=(G, u_i, n ,x)$ which exists a valid $n$-partition such that the disutility of each bundle is no less than $x$ (we call it a $x$-feasible partition),
        Algorithm~\ref{alg:check} returns true. We prove it by induction. For $n=1$, the correctness is obvious. For $n\geq 2$, we select one of the $x$-feasible partition of $I$
        as $P^n_i$ and assume that the algorithm on instance $I$ finds the subtree $T_{c_{j_0}}$.
        It's not hard to observe that one of the following events happens but not both.

        Case 1: $T_{c_{j_0}}$ contains some bundles in $P^n_i$.
        In this case, we know that cutting the subtree $T_{c_{j_0}}$ off the whole $G$ will make the number of left bundles in $P^n_i$ no greater than $n-1$ and the disutility of all left bundles are not decreased
        which means that the instance $(G^{'},u_{i|G^{'}},n-1,x)$ exists a feasible valid $n-1$-partition with respect to $x$. According to our induction hypothesis, the algorithm returns true.

        Case 2: $T_{c_{j_0}}$ is strictly contained in a bundle of $P^n_i$ called $B_1$. We observe that, according to the choice of $c$, $T_c$ has to intersect with at least two bundles $P^n_i$ one of which is $B_1$.
        Due to that fact that $T_{c_{j_0}}$ is strictly contained in $B_1$ which means that $c\in B_1$, there exists a bundle $B_2$ in $P^n_i$ contained in $T_{c_{j_1}}$ where $c_{j_1}\in \mathcal{C}_c$ and $c_{j_1}\neq c_{j_0}$.
        Because $c_{j_0}:=\arg \min_{c_j \in \mathcal{C}_{c}} u_i(T_{c_{j}})$, we know that if $u_i(B_1)\geq x$, $u_i(B_1\setminus T_{c_{j_0}} \cup T_{c_{j_1}}) \geq x$.
        Therefore, we can replace bundle $B_2$ in the remaining instance as bundle $B_1\setminus T_{c_{j_0}} \cup T_{c_{j_1}}$ which implies that there still exists a feasible valid $n-1$-partition with respect to $x$ in the instance $(G^{'},u_{i|G^{'}},n-1,x)$.
        The correctness follows by the induction hypothesis.
    \end{proof}

    We do a binary search and use Algorithm~\ref{alg:check} as a subroutine to compute the MMS value for chores on a tree. The algorithm is presented in Algorithm~\ref{alg:mmstree}.

    \begin{algorithm}[h]
        \caption{MMS-tree($G$, $u_i$, $n$)}
        \label{alg:mmstree}
        \KwIn{An instance $I=(G, u_i, n)$, where $G=(C,E)$ is a tree, $u_i$ is the disutility function of agent $i$ and
         $n$ is the number of bundles.}
        \KwOut{Compute the MMS value for agent $i$ in the instance $I$.}
        $x_l:=0$, $x_u:=u_i(C)$\;
        \While{ $x_l\neq x_u$}
        {
            $x:= \frac{x_l+x_u}{2}$\;
            \If{Check ($G$, $u_i$, $n$, $x$)=true}{$x_l:=x$\;}
            \Else{$x_u:=x$\;}
        }
        \textbf{return} $x$
    \end{algorithm}

Now, we are ready to prove \Cref{MMScom}.
    \begin{proof}
        The correctness of the Algorithm~\ref{alg:mmstree} relies on the correctness of Algorithm~\ref{alg:check} and the definition of the MMS value.
        As for the running time, it is easy to see that there at most $\log(u_i(C))$ iterations, and in each iteration the Check subroutine runs in polynomial time.
        So the Algorithm runs in polynomial times.

        As for cycle, we can enumerate one of the cutting edge $e_i\in E$, then we run MMS-tree($G-e_i$, $u_i$, $n$). The maximum value among all enumerations
        is the MMS value for agent $i$ in the cycle.
    \end{proof}

\section{Allocating Goods on a Cycle to Three Agents}\label{appen-2}
In this section, we consider the problem of allocating goods on a cycle to three agents and prove \Cref{lemgood}.
We will extend our linear programming method for chores in \Cref{cyclechores} to goods.
The extension is intuitive. Almost all statements and arguments only need simple modifications to fit the setting of goods.

\begin{lemma}\label{subsetgg}
        Assume that the instance has three agents and let $\alpha\leq1$. If there is an $\alpha$-MMS$_i$ split $P_i$ of $G$ for agent $i$ and an $\alpha$-MMS$_j$ split $P_j$ of $G$ for agent $j$ such that one bundle in $P_i$ is a subset of one bundle in $P_j$,
    then  $\alpha$-MMS allocations for this instance always exist.
\end{lemma}
\begin{proof}
        Let the agent set be $\{1,2,3\}$ and let $P_i=\{B_{i1},B_{i2},B_{i3}\}$ be an $\alpha$-MMS$_i$ split for agent $i$ ($i\in\{1,2,3\}$).
        Without loss of generality, we assume that the bundle $B_{11}$ in $P_1$ is a subset of the bundle $B_{21}$ in $P_2$ and show
        that there is an $\alpha$-MMS allocation.

        Case 1. $u_3(B_{11})\leq mms_3$: For this case, we have that $u_3(G-B_{11})\geq 2 mms_3$. For any partition of $G-B_{11}$ into two connected bundles, the utility of at least one bundle is not less than $mm_3$ for agent 3.
        The same holds for agent 2. Therefore, we assign $B_{11}$ to agent 1. Then in the instance of the remaining good, we find an MMS split for agent 2 and letting agent 3 get the most valuable bundles for him. The remaining goods are allocated to agent 2.
        It is obviously an $\alpha$-MMS allocation.

        Case 2. $u_3(B_{11})> mms_3$: We assign $B_{11}$ to agent 3. Similar to the discussion in case 1, the MMS value of both agent 1 and agent 2 in the instance of the remaining good does not decrease where MMS allocation for 2 agents always exists.

\end{proof}

    We use the same \emph{cut representation} of the partition notion for chores on cycles in the following lemmas for goods.

    \begin{lemma}\label{goods1}
        Assume that the instance has three agents. Let $P_1=\{e_{11}, e_{12}, e_{13}\}, P_2=\{e_{21}, e_{22}, e_{23}\}$ and $P_3=\{e_{31}, e_{32}, e_{33}\}$ be the cut representations of $\alpha$-MMS splits of $G$ for agents 1, 2 and 3, respectively, which always exist for $\alpha \leq 1$.
        If the instance has no $\alpha$-MMS allocation, then it holds that\\
        (a) all the edges in $P_1, P_2$ and $P_3$ are different;\\
        (b) we can relabel the edges in the three sets $P_1, P_2$ and $P_3$ such that the nine edges appear in the following order
        on the cycle
        $\ \ $

        $G: e_{11}e_{21}e_{31}e_{12}e_{22}e_{32}e_{13}e_{23}e_{33}$
    \end{lemma}
    \begin{proof}
        Equipped with Lemma~\ref{subsetgg}, the proof of the lemma is exactly the same as that of \Cref{lpuse1}.
    \end{proof}

    Let $P_1$, $P_2$ and $P_3$ be the edge representations of an MMS split of $G$ for agents 1, 2 and 3, respectively.
    After deleting the edges in $P_1\cup P_2 \cup P_3$, the cycle $G$ will be split into several connected parts. We call each connected part a \emph{segment}.

    \begin{lemma}\label{goods2}
    Assume that the instance has three agents. Fix an MMS$_i$ split of $G$ for each agent $i$. Let $\alpha\leq 1$.
    If the instance has no $\alpha$-MMS allocation, then for any agent $i$, the utility of any two consecutive segments is less than $\alpha\cdot mms_i$.
    \end{lemma}

    \begin{proof}
        Let $b$ denote two consecutive segments. Assume that $u_{i_0}(b)\geq \alpha \cdot mms_{i_0}$ holds for an agent $i_0\in A$.
        We show that there is always an $\alpha$-MMS allocation.

        Case 1. $b$ is contained in a bundle among the MMS$_{i_0}$ split of $G$ for agent $i_0$: W.L.O.G, we assume $i_0=1$ and $b$ is $\{S_1,S_2\}$. Then there is an $\alpha$-MMS$_{i_0}$ split for agent $i_0$ such that one bundle in it is exactly $b$, i.e. $\{S_1,S_2\},\{S_3,S_4,S_5,S_6 \},\{S_7,S_8,S_9 \}$.
        Note that any MMS split is also an $\alpha$-MMS split for $\alpha \leq 1$.
        So the condition of \Cref{subsetcc} holds and then $\alpha$-MMS allocations always exist.

        Case 2. $b$ is not contained in a bundle among the MMS$_{i_0}$ split of $G$ for agent $i_0$. For this case, it is easy to see that the remaining path $G-b$ consists of two connected paths, each of which is a supset of a bundle
        in the MMS split of $G$ for an agent other than $i_0$. We assign these two parts to these two agents and assign $b$ to agent $i_0$. This will be an $\alpha$-MMS allocation.

        In any case, we can find an $\alpha$-MMS allocation, a contradiction. So the lemma holds.
    \end{proof}

    \begin{lemma}\label{goods3}
        Assume that the instance has three agents. Let $\alpha\leq 1$.
         If the instance has no $\alpha$-MMS allocation, then for any $\alpha$-MMS$_i$ split $P_i$ of $G$ for agent $i\in A$, there is one bundle $b$ in $P_i$ such that $u_j(b)\geq \alpha \cdot mms_j$ holds for any agent $j\in A\setminus \{i\}$, and for all other bundles $b'$ in $P_i$ it holds that $u_j(b')< \alpha \cdot mms_j$ for any agent $j\in A\setminus \{i\}$.

    \end{lemma}
\begin{proof}
    We say that an agent is satisfied with a bundle if the utility of this bundle is at least $\alpha$ times of his MMS value.
    We know that agent $i$ is satisfied with all the three bundles in $P_i$.
    For the other two agents $\{j_1,j_2\}=A\setminus \{i\}$, each is satisfied with at least one bundle in $P_i$.
    If agent $j_1$ and agent $j_2$ are satisfied with two different bundles in $P_i$, then we assign these two bundles in $P_i$ to them and assign the last bundle in $P_i$ to agent $i$.
    This would be an $\alpha$-MMS allocation. So we know that agent $j_1$ and agent $j_2$ are satisfied with the same bundle in $P_i$.
\end{proof}

    We are ready to describe the linear programming model. Here, we continue to use the same variables notations as chores and
    we declare the new constraints in the goods situations.

    The first set of constraints in the LP model is to set the domain of these variables, i.e.
    \begin{eqnarray}\label{cequs1}
        x_j,y_j,z_j\geq 0 ~~~~~~(j=1,2,\dots,9).
    \end{eqnarray}

    We normalize the MMS value for each agent by letting  $mms_1=mms_2=mms_3=1$. According to the MMS splits of the three agents, we get the following $3\times 3$ constraints.

    \begin{eqnarray}
        x_{1+i}+x_{2+i}+x_{3+i}\geq 1&(i=0,3,6);\label{cequs2}\\
        y_{2+i}+y_{3+i}+y_{4+i}\geq 1&(i=0,3,6);\label{cequs3}\\
        z_{3+i}+z_{4+i}+z_{5+i}\geq 1&(i=0,3,6),\label{cequs4}
    \end{eqnarray}
    where the indices are computed modulo 9. We will always assume this without stating it every time.

    By Lemma~\ref{goods2}, we directly get the following $3\times 9$ constraints,
    \begin{eqnarray}
        x_{1+i}+x_{2+i}< \alpha&(i=0,1,\dots,8);\label{cequs5}\\
        y_{1+i}+y_{2+i}< \alpha&(i=0,1,\dots,8);\label{cequs6}\\
        z_{1+i}+z_{2+i}< \alpha&(i=0,1,\dots,8).\label{cequs7}
    \end{eqnarray}

    Next, we consider the constraints generated by Lemma~\ref{goods3} which is similar to chores situation except for the following slightly modification.
    \begin{eqnarray}
        u_i(B_{12})< \alpha \times mms_i =\alpha&(i=2,3);\label{ceqd1}\\
        u_i(B_{13})< \alpha \times mms_i =\alpha&(i=2,3). \label{ceqd2}
    \end{eqnarray}
    Then, we get five LP models as before. By solving the five LP, we find out that the worst ratio is ${\frac{5}{6}}$.
    This means for any $\alpha \leq {\frac{5}{6}}$, our LP will have no solution. We get the following result.


    \begin{lemma}
        $\frac{5}{6}$-MMS allocations of goods on a cycle to three agents always exist and can be found in polynomial time.
    \end{lemma}

    \begin{proof}
        For any instance, we fix an MMS split $P_i$ for each agent $i$ (considering them as the cut representations).
        We let $P_1=\{e_{11}, e_{12}, e_{13}\}$, $ P_2=\{e_{21}, e_{22}, e_{23}\}$ and $P_3=\{e_{31}, e_{32}, e_{33}\}$. Without loss of generality, we assume that they are nine different edges and
        we can relabel them in the order $e_{11}e_{21}e_{31}e_{12}e_{22}e_{32}e_{13}e_{23}e_{33}$.
        (Otherwise, we have a polynomial time algorithm to find a MMS allocation similar to chores situations.)

        According to a fixed MMS split, we instantiate the variables of the LP model with the corresponding value. The LP does not have any solution for $\alpha \geq {\frac{5}{6}}$, which means some constraints in the LP will not hold for $\alpha \geq {\frac{5}{6}}$.
        The constraints (\ref{cequs1}) to (\ref{cequs4}) clearly hold. If some constraints in (\ref{cequs5}) to (\ref{cequs7}) do not hold, then there is an $\frac{5}{6}$-MMS allocation with one bundle containing two consecutive segments.
        Then we can find the $\frac{5}{6}$-MMS allocation in polynomial time by the algorithm stated in the proof of the Lemma~\ref{goods2}. Otherwise, one of (\ref{ceqd1}) and (\ref{ceqd2}) and the following constraints will not hold.
        For this case, one of $P_1$, $P_2$ and $P_3$ is a $\frac{5}{6}$-MMS allocation by Lemma~\ref{goods3} and we can also check it in polynomial time.

        In any case, there is a $\frac{5}{6}$-MMS allocation and it can be found in polynomial time.
    \end{proof}

    The tight instance we get by the LP method is shown in Table~\ref{t2-ex}.
    \setcounter{table}{1}
    \begin{table}[h]
        \centering
        \begin{tabular}{p{40pt}|p{12pt}|p{12pt}|p{12pt}|p{12pt}|p{12pt}|p{12pt}|p{12pt}|p{12pt}|p{12pt}}
            \hline
             &$c_1$&$c_2$&$c_3$&$c_4$&$c_5$&$c_6$&$c_7$&$c_8$&$c_9$\\
            \hline
            Agent 1&$\frac{2}{3}$&$0$&$\frac{1}{3}$&$\frac{1}{3}$&$0$&$\frac{2}{3}$&$\frac{1}{6}$&$\frac{2}{3}$&$\frac{1}{6}$\\
            \hline
            Agent 2&$\frac{1}{2}$&$\frac{1}{3}$&$\frac{1}{2}$&$\frac{1}{6}$&$\frac{1}{6}$&$\frac{1}{2}$&$\frac{1}{3}$&$\frac{1}{2}$&$0$\\
            \hline
            Agent 3&$\frac{2}{3}$&$\frac{1}{6}$&$\frac{2}{3}$&$0$&$\frac{1}{3}$&$\frac{1}{3}$&$0$&$\frac{2}{3}$&$\frac{1}{6}$\\
            \hline
        \end{tabular}
        \caption{An example of nonexistence of $\alpha$-MMS allocations of a 9-cycle to three agents for any $\alpha >{\frac{5}{6}}$}\label{t2-ex}
    \end{table}

\end{document}